\documentclass{iaus}
\usepackage{graphicx}

\title[Active Cores in Deep Fields]
{Active Cores in Deep Fields}
\author[M\"uller \& Hasinger]
{Andreas M\"uller$^1$ \and G\"unther Hasinger$^2$}
\affiliation{Max--Planck--Institute for Extraterrestrial Physics, p.o. box 1312, D--85741 Garching, Germany \break $^1$email: amueller@mpe.mpg.de
\ $^2$email: ghasinger@mpe.mpg.de}

\pubyear{2005}
\volume{230}  
\pagerange{1--8} 
\date{September 2005}
\jname{Populations of High Energy Sources in Galaxies}
\editors{Evert J.A. Meurs \& Giuseppina Fabbiano, eds.}

\begin{document}

\maketitle

\begin{abstract}
Deep field observations are an essential tool to probe the cosmological evolution 
of galaxies. In this context, X-ray deep fields provide information about some of the most energetic 
cosmological objects: active galactic nuclei (AGN). Astronomers are interested in detecting sufficient 
numbers of AGN to probe the accretion history at high redshift. This talk gives an overview of the 
knowledge resulting from a highly complete soft X--ray selected sample collected with \textit{ROSAT}, 
\textit{XMM--Newton} and \textit{Chandra} deep fields. The principal outcome based on X--ray luminosity 
functions and space density evolution studies is that low--luminosity AGN evolve in a dramatically 
different way from high--luminosity AGN: The most luminous quasars perform at significantly earlier
cosmic times and are most numerous in a unit volume at cosmological redshift $z\sim 2$. In contrast, 
low--luminosity AGN evolve later and their space density peaks at $z\sim 0.7$. This finding is also 
interpreted as an anti--hierarchical growth of supermassive black holes in the Universe. Comparing 
this with star formation rate history studies one concludes that supermassive black holes enter the 
cosmic stage before the bulk of the first stars. Therefore, first solutions of the so--called hen--egg 
problem are suggested. Finally, status developments and expectations of ongoing and future extended 
observations such as the \textit{XMM--COSMOS} project are highlighted.
\keywords{surveys, galaxies: active, galaxies: evolution, galaxies: high--redshift, 
galaxies: luminosity function, mass function, galaxies: nuclei, galaxies: statistics, 
X--rays: diffuse background, X--rays: galaxies, X--rays: general}
\end{abstract}

\firstsection 
%
%
\section{Introduction}
Deep X--ray surveys turned out to be valuable diagnostic tools to investigate the galaxy formation 
history and large scale structure of the Universe. The diffuse X--ray background (XRB) at the sky 
is composed of discrete sources that have been almost resolved by deep \textit{ROSAT}, \textit{ASCA}, 
\textit{Chandra} and \textit{XMM--Newton} observations at photon energies in the 0.1-10 keV band 
\footnote{However, it must be stated that at higher photon energies in the 5-10 keV band there is 
still a lack of identification.} 
(\cite{Hasinger98, Mushotzky00, Giacconi01, Hasinger01, Giacconi02, Alexander03, Worsley04, Bauer04}).
The X--ray sources in deep field observations are mostly active galactic nuclei (AGN). This has been
confirmed by optical identification programmes with the \textit{Keck} (\cite{Schmidt98, Lehmann01, Barger01, 
Barger03}), the \textit{VLT} (\cite{Fiore03, Szokoly04, Zheng04, Mainieri05}) and \textit{COMBO--17} 
(\cite{Wolf04}). All AGN are captured in different evolutionary stages. Therefore, X--ray deep fields 
also probe the \textit{in vivo} growth of supermassive black holes (SMBHs) that drive the AGN luminosities 
according to a widely accepted paradigm.
Optical identification programmes provide cosmological redshifts of the X--ray sources. This can be 
done by spectroscopic or, especially at high redshifts, photometric methods (\cite{Zheng04, Mainieri05, 
Wolf04}). Optical observations demonstrate that the number distributions in redshift space of 
all galaxy types peak around $z\simeq 0.7$. Further, the AGN sample can be classified into AGN type--1
(unabsorbed, unobscured) and AGN type--2 (absorbed, obscured) by using optical and/or X--ray methods.
In the optical, AGN type--1 are defined as sources exhibiting broad Balmer emission lines due to the 
fact that the observer is able to view the core of the AGN from the outside. These features are lacking 
for AGN type--2. In X--rays, AGN type--1 show an unabsorbed X--ray spectrum whereas AGN type--2 have
absorbed spectra around photon energies from 0.5-1 keV. \\
The paper is organised as follows: In Sec. \ref{sec:samp} the soft X--ray selected sample is presented,
Sec. \ref{sec:xlf} treats the method to analyse features and evolution of different AGN object classes 
by means of X-ray luminosity functions. Space density and luminosity density evolutions as other tools 
are presented in Sec. \ref{sec:dens}. The observations are interpreted as a scenario described in 
Sec. \ref{sec:grow}. In Sec. \ref{sec:discuss} the results are compared to observational data from other 
work. Finally, we conclude in Sec. \ref{sec:concl}. 

%
%
\section{The X--ray selected AGN type--1 sample} \label{sec:samp}
A sample of about 1000 AGN type--1 is considered that has been merged from \textit{ROSAT}, 
\textit{Chandra} and \textit{XMM--Newton} surveys. The sources are restricted to the 
0.5--2 keV band and cover five orders of magnitude in flux and six orders of magnitude 
in survey solid angle. This sample is with 95\% highly complete allowing to construct luminosity 
functions over cosmological timescales with unprecedented accuracy. Details of the soft 
X--ray sample are presented in \cite{Hasinger04} (H04 hereafter; see Table 1 therein).
The advantage of restricting the sample only to type--1 AGN is that systematic
uncertainties are excluded \textit{a priori}. AGN type--2 can introduce these uncertainties
by the varying and typically unknown AGN absorption column densities.
As demonstrated by \cite{Szokoly04} the optical AGN classification scheme suffers from
dilution effects of AGN excess light from stars in the host galaxy -- especially at low
X--ray luminosities and medium redshifts. Then, only X--ray AGN classification schemes
as introduced by \cite{Schmidt98} and \cite{Szokoly04} can do the job.
As anticipated AGN are the main contributors to the XRB accounting for a fraction of about
70--100\%. A review about AGN and their engines, SMBHs, can be found in the PhD thesis by 
\cite{Mueller04}. It was found that at faintest X--ray fluxes there are contributions by 
other population classes such as starburst and normal galaxies, see 
\cite{Hornschemeier00, Hornschemeier01}. 

%
%
\section{X-ray luminosity functions} \label{sec:xlf}
\begin{figure}
\includegraphics[width=9.0cm]{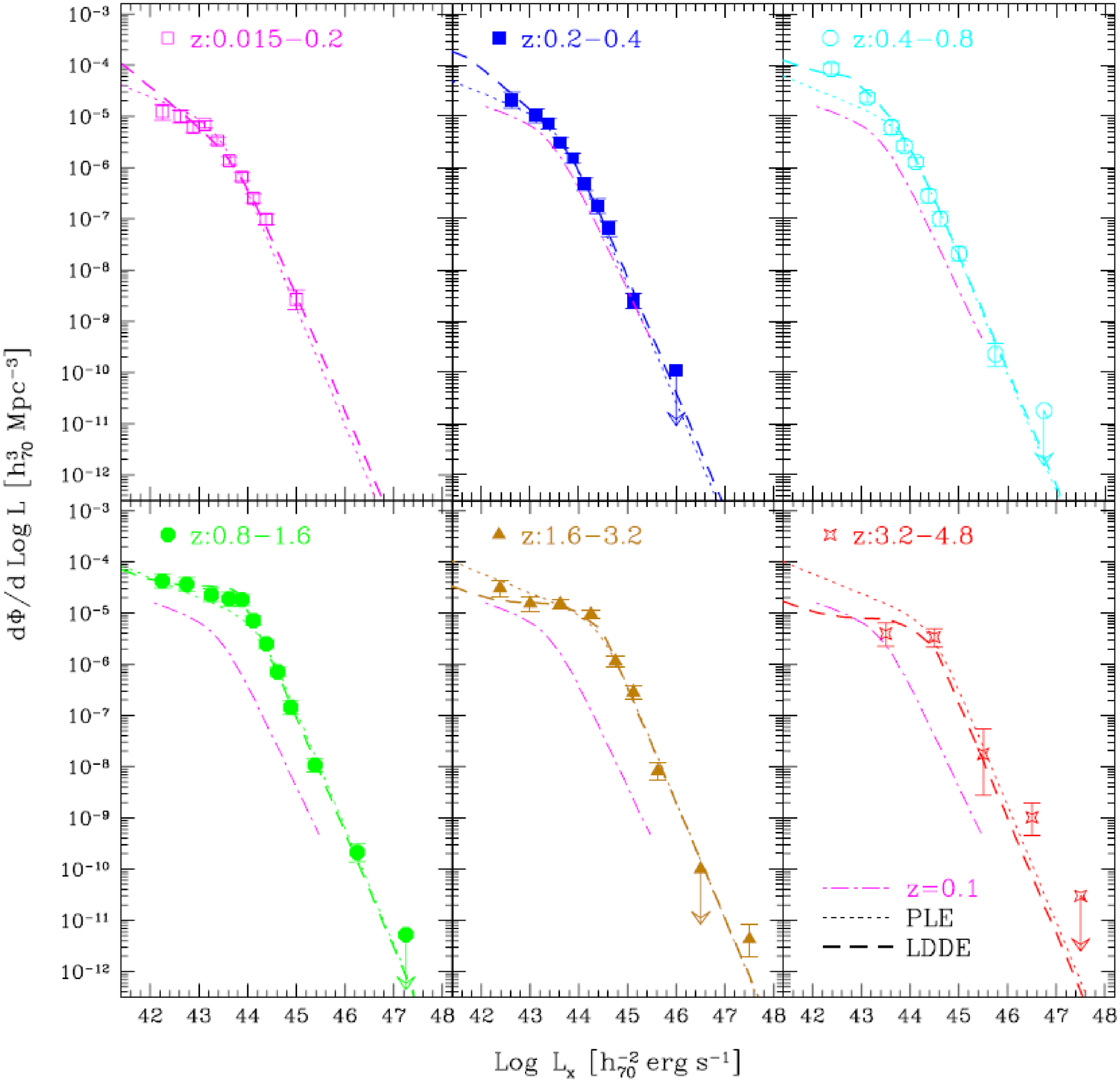}
\caption{The soft X--ray luminosity function of the soft X--ray selected AGN type--1 sample in 
different redshift shells as denoted. Error bars correspond to 68\% Poisson errors of the
AGN number in the bin. The best--fit two power--law model in the lowest $z$--shell 
$z=0.015-0.2$ is overplotted in each panel for reference. Dotted and dashed lines give best--fit
PLE and LDDE models.} \label{fig:sxlf}
\end{figure}

As elaborated in \cite{Hasinger05} (HMS05 hereafter) the X-ray luminosity functions (XLFs) of the 
sample presented in Sec. \ref{sec:samp} are deduced by following two methods: The first ('binned') 
method is presented in \cite{Miyaji00} and is based on a variant of the $1/V_{a}$ method. In 
this approach, the binned luminosity function in a given redshift bin $z_i$ is derived by 
dividing the observed number $N_\mathrm{obs}(L_x, z_i)$ by the corresponding volume to the 
redshift range and appropriate survey X--ray flux limits and solid angles. Each of the
XLFs is fitted by an analytical function to determine the bias in this luminosity value 
caused by a gradient of the XLF across one bin. By means of the resulting analytical function
one can predict $N_\mathrm{mod}(L_x, z_i)$. The ratio $N_\mathrm{obs}/N_\mathrm{mod}$ 
serves as a factor to correct the XLF due to bias to first order. \\
The second ('unbinned') method is based on unbinned data. Individual $V_\mathrm{max}$ values 
from \textit{ROSAT Bright survey} (RBS) sources are used to evaluate the zero--redshift luminosity 
function (0zLF). By means of this LF the number of derived RBS sources matches accurately 
the observed number. Hence, the advantage of this method is that it is free from bias effects 
emerging in the first method. Evolution, i.e. space density as function of redshift, comes 
into play by introducing binning in luminosity and redshift. Again, bias effects are avoided 
by iterating the parameters of an analytical representation of the space density function. 
Together with the 0zLF this is used to predict $N_\mathrm{mod}(L_x, z_i)$ for the surveys. 
Finally, the observed space densities in each bin are evaluated by multiplying by the ratio 
$N_\mathrm{obs}/N_\mathrm{mod}$ with the space density value.
The result is shown in Fig. \ref{fig:sxlf}: This is the soft X--ray luminosity function (SXLF)
in different redshift shells ranging from $z=0.015$ to $z=4.8$ of about 1000 AGN type--1. 
It clearly demonstrates that the LF shape varies with redshift. The typical two power--law model 
is confirmed. Fits of the $z$--dependent LF profiles strongly suggest a luminosity--dependent 
density evolution (LDDE) as outlined in HMS05. Pure luminosity evolution (PLE) 
overpredicts especially the LF behavior at high redshift, $z\sim 2-5$.
%
%

%
%
\section{Space density and luminosity density evolution} \label{sec:dens}
\begin{figure}
\includegraphics[width=9.0cm]{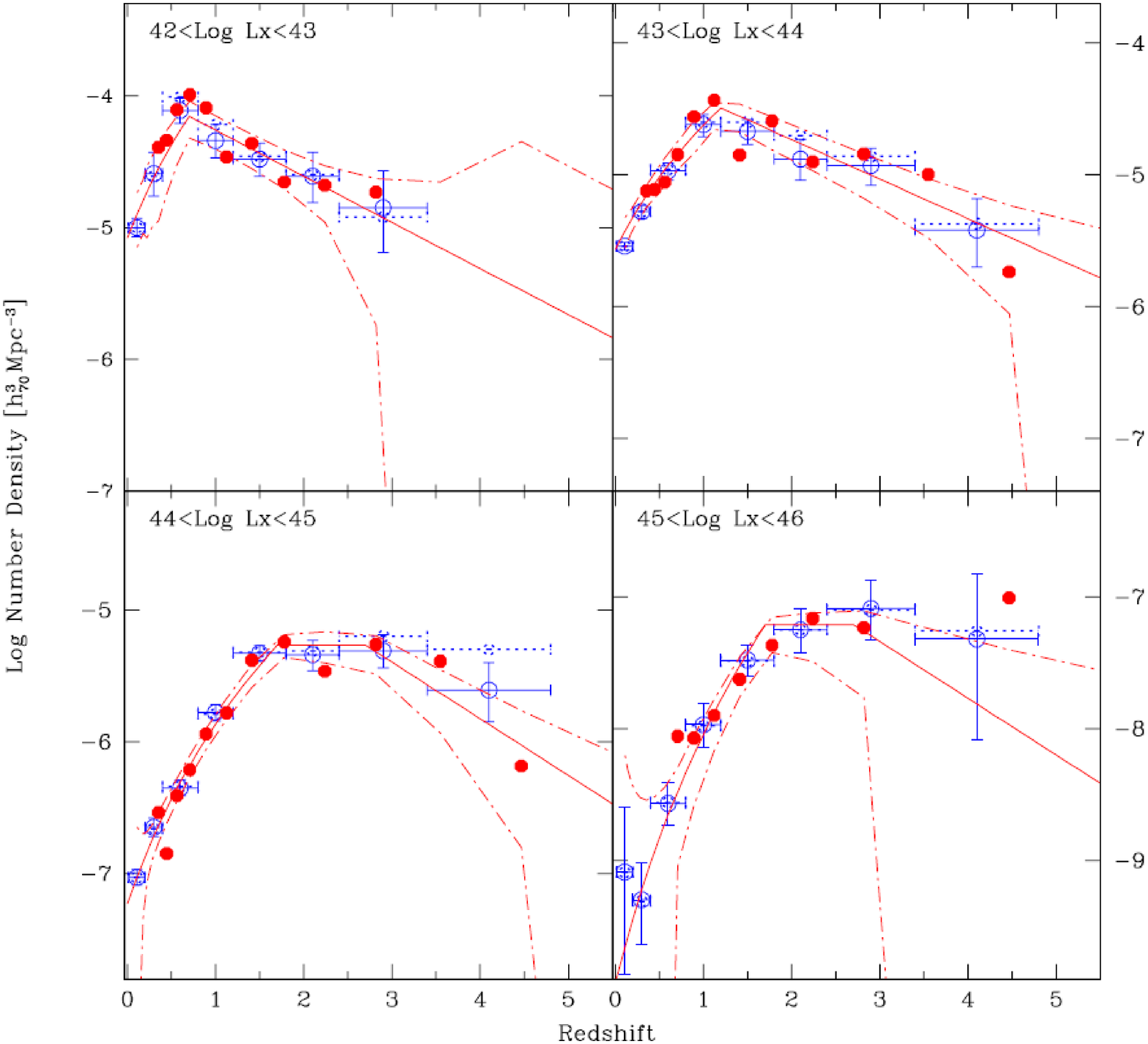}
\caption{Space density evolution in different luminosity classes as derived from the binned 
(blue) and unbinnend method (red) outlined in Sec. \ref{sec:xlf}. Dashed horizontal blue lines
correspond to the maximum contribution of unidentified sources.} \label{fig:sp-dens-ev}
\end{figure}

An alternative plot to analyse the data is shown in Fig. \ref{fig:sp-dens-ev}: data are 
divided into luminosity classes and plotted over redshift. The plot clearly demonstrates the 
space density evolution for each luminosity class, i.e. compares low--luminosity AGN (LLAGN) to 
high--luminosity AGN (HLAGN).The essential and surprising result is that the space density of
HLAGN peaks at significantly higher redshift, $z\sim 2$, i.e. earlier cosmic times, than the 
LLAGN that peak at $z\sim 0.7$. In other words: Luminous quasars formed first and the bulk of 
faint AGN such as low--luminosity Seyferts came significantly later on the cosmic stage.
%
%
Another remarkable fact is that for the first time the high--redshift decline in all luminosity 
classes with $\log\left(L_X/[\mathrm{erg}/\mathrm{s}]\right)\lesssim 45$ is shown. At higher 
luminositites insufficient statistics does not allow to derive secure trends.
Space density evolution can also be studied using optically selected AGN type--1 as demonstrated
by \cite{Croom04} and \cite{Fan01} using data sets from the 2dF and 6dF QSO redshift surveys. A 
comparison of an optical (UV) vs. a X--ray selection of AGN type--1 is elaborated in H04. The main 
result is that there is relatively little difference in space density evolutions between the two 
wave bands; however, we have to consider that (as demonstrated in Fig. \ref{fig:sp-dens-ev})
rise and fall of the space density is \textit{X--ray luminosity dependent}. Therefore, comparisons
of this kind are only preliminary and have to be re--analysed when larger samples of high--redshift 
X--ray selected QSOs are available. \\
A fruitful comparison is based on analysing flux correlations between optical and X--ray fluxes.
The X--ray data from H04 and optical SDSS data from \cite{Vignali03} are considered.
Then, the X--ray flux in the soft 0.5-2 keV band is plotted over optical $\mathrm{AB_{2500}}$ 
magnitudes. The first result is that the X--ray multi--cone survey covers a significantly wider
range. Due to these better preconditions, a tight correlation was discovered by H04.
Plotting the source's luminosity in the same filters indicate an essential difference: the luminosities
in X--rays vs. those in redshifted 2500 $\mathring{\mathrm{A}}$ \textit{scale linearly}, $L_\mathrm{X}\propto L_{\mathrm{UV}}$,
when using X--ray selected data. But the correlation is non--linear, $L_\mathrm{X}\propto L^{0.75}_{\mathrm{UV}}$,
when UV data are used. A preliminary interpretation of this discrepancy are sample selection effects. 
Host galaxy contamination corrections that are missing in the X--ray analysis are likely to be not 
responsible for the difference as $\alpha_{ox}$ value studies indicate (for details see H04).

%
%
\section{Growth of supermassive black holes} \label{sec:grow}
The analysis of X-ray deep field data by using XLFs and luminosity--weighted space density evolutions 
provide new clues for our understanding of AGN evolution in general and black hole growth in particular.
AGN luminosity is connected to the central SMBH via Eddington's relation. Hence, from the observed 
activity the black hole mass can be estimated. Concerning HLAGN the soft X--ray selected samples clearly 
trace the rise and fall of quasars. The most luminous objects emerge firstly, grow rapidly by accretion 
and survive as supermassive dark remnants in the local Universe. Today, we observe these SMBHs in the 
centers of massive galaxy clusters such as M87 in the Virgo cluster. This scenario gets support from 
theoretical simulations presented recently by \cite{Springel05}: In hydrodynamical simulations of structure 
formation on large scales performed in a cube with $\approx$ 700 Mpc size, the evolution of cold dark matter 
(CDM) halos are followed from $z\sim 120$ to $z\sim 0$. It has been shown that the observed large scale structure 
with superclusters surrounded by smaller clusters can be reproduced. Furthermore, the simulations strongly 
suggest that the most massive black holes with $10^{10} \ \mathrm{M}_\odot$ \footnote{and also the oldest 
population of stars (PopIII)} can be found in the centers of the most massive clusters. These are the 
descendants of the most active accretors in the quasar era at $z\sim 2$. X-ray deep field data from the soft 
band now suggest an \textit{anti--hierarchical growth} of black holes: the most massive black holes with 
$10^8-10^{10} \ \mathrm{M_\odot}$ formed first and the bulk of the lighter black holes with 
$10^6-10^8 \ \mathrm{M_\odot}$ formed later. This behaviour is totally unexpected and not predicted by 
standard CDM structure formation scenarios which are hierarchical. Maybe this observational fact hints for 
two accretion modes that differ in accretion efficiency (\cite{Duschl02}). However, a self--consistent model 
which explains both, anti--hierarchical black hole growth and the local black hole mass function derived from 
the $M_\mathrm{BH}-\sigma$ relation assuming two accretion modes has been suggested recently by \cite{Merloni04}. 
Concretely, two accretion modes can be established by a varying value of efficiency $\epsilon$ -- the parameter 
in accretion theory that controls conversion of mass flux into radiation flux. Black hole angular momentum controles 
crucially the efficiency: it is rather low, $\epsilon\sim 0.054$, for Schwarzschild black holes, or high, 
$\epsilon\sim 0.37$, for Kerr black holes as has been found by \cite{Thorne74}.
The SXLFs that give rise to an early quasar era in combination with the evolution of the star formation 
rate (SFR) also provide new clues to the so--called \textit{hen--egg problem} i.e. if galaxies or the 
SMBHs came first in the Universe. The history of global cosmic star formation has been determined using 
Hubble deep field (HDF) observations (see \cite{Madau96, Connolly97}) corrected for dust obscuration 
following \cite{Pettini97}. The resulting SFR as function of redshift shows only a moderate 
variation with cosmic time with a possible peak around a redshift of unity and a steep decline towards 
lower redshifts. But the AGN population reveals a pronounced peak at significantly higher redshifts
as shown in the SXLF of AGN type--1. This discrepancy indicates that the bulk of SMBHs have been in 
place \textit{before} the bulk of stars in these galaxies formed. \\ 
Very recently, an analysis of the specific SFR (SSFR) i.e. the star formation rate per unit stellar mass,
has been presented based on FORS Deep Field (FDF) and GOODS--S field data, see \cite{Feulner05}. Their 
main finding is that the most massive galaxies with stellar masses around $10^{11} \ \mathrm{M_\odot}$ 
are in quiescent state for $z\lesssim 2$ but that the SSFR is increased by a factor of $\sim$ 10
for redshifts $z\gtrsim 2$. This is interpreted as a very early formation epoch of most massive galaxies 
as supported by the Millennium simulation (\cite{Springel05}) and X--ray deep field observations 
(HMS05). Comparisons of deep field observations therefore allow for attractive solutions 
of the hen--egg problem.

%
%
\section{Discussion} \label{sec:discuss}
In this section the analysis with soft X-ray data of about 1000 AGN type--1 are compared to 
other work. \cite{Ueda03a} analysed the hard X--ray luminosity functions (HXLFs) of bright AGN using 
\textit{ASCA} data. 
This hard X--ray selected sample is highly complete ($\sim$ 95\%) and consists of both, AGN type--1
and type--2, comprising $\approx$ 230 sources. The main feature is that the fraction of type--2 AGN 
decreases with intrinsic luminosity or in other words: the space density of obscured AGN is 
luminosity--dependent. A possible explanation for this trend may be due to clean--out effects 
because the luminous AGN core may blow away the dusty torus at the pc--scale by radiation. 
Low--luminosity cores are not strong enough to initiate such a decay of the mass reservoir. 
Therefore, the AGN classification scheme into type--1 and type--2 is not only a pure orientation 
effect but is also dependent on AGN luminosity. As outlined in Sec. \ref{sec:xlf} the SXLFs strongly 
suggest a LDDE. Interestingly, the hard X--ray selected sample by \cite{Ueda03b} demonstrates the same 
behaviour. There is only a difference in the normalization by a factor of five -- probably due to absorbed 
objects missing in the SXLF. Another soft and hard X--ray selected sample based on \textit{Chandra} 
(\textit{CDF--N, CDF--S, CLASXS}) and \textit{ASCA} data has been presented recently by \cite{Barger05}. 
Their results are in good agreement with the SXLF analysis presented here, but they still suffer from 
substantial identification incompleteness. Their AGN type--1 sample does not directly compare to the one 
discussed here because \cite{Barger05} only include optically classified AGN type--1 (broad--line AGN) but 
miss most of LLAGN type--1 that are considered here. A critical comparison of all available XLFs from different 
observations will be the topic of an upcoming paper. \\
Recently it turned out that there is a possible new population of star--forming galaxies that can be found 
especially at very low fluxes, $S_\mathrm{X}\lesssim 10^{-16} \ \mathrm{erg}\,\mathrm{cm}^{-2}\,\mathrm{s}^{-1}$ 
as discovered by \cite{Hornschemeier00, Hornschemeier01, Rosati02, Norman04}. Upcoming X--ray selected samples 
will have to account for this special class by lowering the flux limits. 

%
%
\section{Conclusions} \label{sec:concl}
In general, X--ray deep fields prove valuable tools to investigate the formation and evolution history
of galaxies, in particular of AGN.
Concerning AGN, X--ray data sets cannot stand alone; they need additional support from optical 
identification programmes that discriminate type--1 vs. type--2 and deliver redshift
determinations by spectroscopic or photometric techniques. Then, the analysis follows two 
branches: samples are constructed from soft and/or hard X--ray energy bands. To date soft X--ray 
selected samples provide high--quality data samples with high degrees of completeness. Hard X--ray
selected samples still suffer from lacking spectroscopic identification. There is much work to be done 
in the future to resolve the high--energy XRB branch into discrete sources. Nevertheless, the comparison
of the results from soft (HMS05) and soft plus hard X--ray selected samples by \cite{Ueda03a} and 
\cite{Barger05} agree well so far. \\
The principal method to analyse X--ray samples are based on deriving XLFs, space density evolution and 
luminosity density evolution. The SXLF sample presented here strongly suggests dramatically different 
evolutionary paths of LLAGN and HLAGN: Most luminous quasars formed significantly earlier as the space 
density peak at $z\approx 2$ demonstrates. The bulk of galaxies with low luminosities such as Seyferts 
perform later with a space density peak at $z\approx 0.7$. Linking AGN luminosiy to black hole mass via 
the Eddington criterion, one immediately arrives at the finding that the growth of SMBHs is anti--hierarchical: 
The most massive holes form first. A comparison with the SFR evolution even shows that the most massive 
holes formed before the bulk of first stars formed. Hence, deep field observation have the power to solve 
the cosmological hen--egg problem. The observed AGN evolution scenario should gain support from theoretical 
simulations. The Millennium simulation is one essential cornerstone that allows to predict the evolution of 
the large scale structure. \\
%
\begin{figure}
\includegraphics[width=6.0cm]{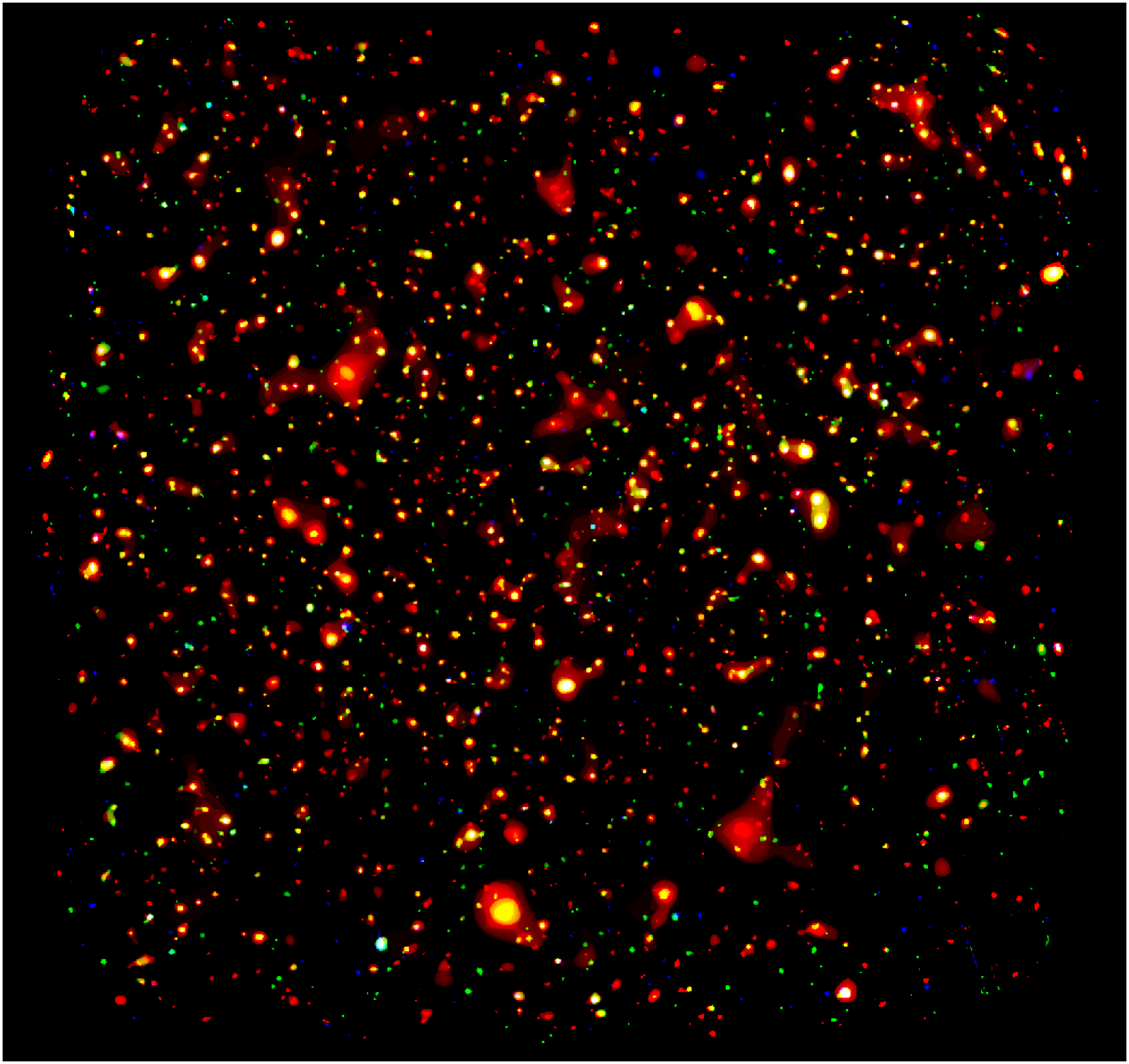}
\caption{RGB image of the \textit{XMM--COSMOS} deep field first year data. The 2 square degree field 
consists of 25 separate pointings amounting 1.4 Ms exposure time in total. Red colors belong to 
0.5-2 keV photon energies, green to 2-4.5 keV and blue to 4.5-10 keV. \textit{Courtesy:} Nico 
Cappelluti and Alexis Finoguenov (both MPE Garching)} \label{fig:xmmcosmos}
\end{figure}
%
The next observational steps are to collect better data i.e. identify more sources, perform deeper pencil 
beams and adjust suitable flux limits. One upcoming multi--wavelength project is called \textit{COSMOS}. 
This is a \textit{HST} treasury project where a 2 square degree field is observed with \textit{ACS} by 
using 10\% of observing time over two years. There are commitments from \textit{VLA, VLT, Subaru} and 
\textit{XMM--Newton} so that there will be also radio to X--ray data available for the same field where
more than two million sources are supposed to be detected. Main goals of \textit{COSMOS} are to study the 
large scale structure (LSS), evolution of galaxies, AGN and dark matter haloes, to investigate the SFR, 
and to derive AGN activity as function of morphology, size, redshift and LSS environment.
One pencil beam with 0.2 square degree is included. The sensitivity level amounts $5\times 10^{-16} \ \mathrm{erg}\,\mathrm{cm}^{-2}\,\mathrm{s}^{-1}$ and captures therefore sources that are by a factor of 
two fainter than in deepest \textit{ROSAT} surveys \footnote{Detailed information is provided on the 
\textit{COSMOS} website \texttt{http://www.astro.caltech.edu/$\sim$cosmos/} and the \textit{XMM--COSMOS} 
website \texttt{http://www.mpe.mpg.de/XMMCosmos}}.
First year observations of \textit{XMM--COSMOS} started in December 2003. AO3 data are now available: 
A preliminary RGB image of 2 square degree size and 1.4 Ms total exposure time is shown in Fig. 
\ref{fig:xmmcosmos}. As expected, a wealth of sources, almost AGN, can be seen. The analysis of these 
\textit{XMM--COSMOS} data are subject of a future paper (\cite{Hasinger06}). Other upcoming extended 
deep X-ray surveys such as \textit{eRosita} and \textit{eCDFS} will improve significantly the analysis 
presented here. 

%
%
\begin{acknowledgments}
AM and GH would like to thank the organizers of the IAU symposium No. 230 in Dublin, Ireland.
\end{acknowledgments}

%
%

\end{document}